# NEW RESULTS ON THE ANOMALOUS PRECURSOR DIAMAGNETISM IN THE UNDERDOPED $La_{1.9}Sr_{0.1}CuO_4$ SUPERCONDUCTOR.

J. D. Dancausa, N. Cotón, J. M. Doval, J. Mosqueira, M. V. Ramallo, A. Ramos-Álvarez, R. I. Rey and F. Vidal[*]

LBTS, Universidade de Santiago de Compostela, ES-15782, SPAIN

*felix.vidal@usc.es, Tel.: +34-881-814031, Fax: +34-881-814 112

**Abstract**

The new results summarized here, including a brief comparison with the paraconductivity, further suggest that the anomalous precursor (above $T_c$) diamagnetism recently observed in the underdoped $La_{1.9}Sr_{0.1}CuO_4$ superconductor could be attributed to the presence, in addition to the conventional superconducting pair fluctuations, of $T_c$-inhomogeneities with long characteristic lengths associated with chemical disorder .

## 1. Introduction.

In high-$T_c$ cuprate superconductors (HTSC), the interplay between normal state properties, unconventional superconductivity and superconducting fluctuations is a central and still open debate [1]. A particular aspect of these interplays, but whose clarification may contribute to disentangle this debate, is the long standing dilemma between seemingly anomalous superconducting fluctuations, beyond the phenomenological Gaussian-Ginzburg-Landau (GGL) scenario, and critical temperature inhomogeneities [2]. In the case of the in-plane fluctuation diamagnetism above $T_c$ (the so-called *precursor diamagnetism*) in underdoped cuprates, $\Delta\chi_{ab}$, the anomalies observed by various research groups were easily explained by the presence, in addition to the conventional (GGL) superconducting fluctuations, of $T_c$-inhomogeneities with long characteristic lengths [larger than the in-plane superconducting coherence length amplitude, $\xi_{ab}(0)$], associated with chemical disorder [3, 4]. However, the majority of the recent works on that issue, including the most influential, again support anomalous superconducting fluctuations (non- GGL) in these compounds (see, e. g., Refs. 5 to 7, and references therein). In contrast, our last results on $\Delta\chi_{ab}$ in two samples of the underdoped La$_{1.9}$Sr$_{0.1}$CuO$_4$, with different chemical disorder, further suggest that these seemingly anomalies may in fact be attributed to the presence (unavoidable in non-stequiometric compounds) of $T_c$-inhomogeneities [8].

We will devote this paper to clarify two aspects not included in Ref. 8: i) The direct interrelation between some of the isothermal magnetization anomalies observed near T$_c$ in the low field regime and the magnetic transition width. ii) The relationship between the $\Delta\chi_{ab}$ data of Ref. 8 and the in-plane paraconductivity, $\Delta\sigma_{ab}$, earlier measured by Currás and coworkers in La$_{1.9}$Sr$_{0.1}$CuO$_4$ films [9]. The interest of this last aspect is enhanced by recent results suggesting an unconventional relationship between the fluctuation effects on the conductivity in the terahertz range and $\Delta\chi_{ab}$ [10-12], although the (dc) paraconductivity is not directly comparable with these interesting measurements of Bilbro and coworkers [10,11].

## 2. Inside the full inhomogeneous region: Precursor diamagnetism anomalies and magnetic transition width.

The temperature and magnetic regions around the average critical temperature, $\bar{T}_c$, where the $T_c$-inhomogeneities may deeply affect the precursor diamagnetism measurements, may be illustrated schematically as indicated in Fig. 1. The dashed areas represent the broadening of the $H_{c1}(T)$ and $H_{c2}(T)$ lines due to the $T_c$ distribution, that here is supposed to be Gaussian. These dashed areas roughly represent an *H-T* region that we have called "*full inhomogeneous region*" [8], and the corresponding reduced-temperature width above $\bar{T}_c$ may be estimated as,

$$\varepsilon^{inh}(H) = 2(\Delta T_c/\bar{T}_c) - H/H_{c2}(0), \tag{1}$$

where $\Delta T_c$ is the magnetic transition width that in turn may be estimated from the temperature dependence of the field-cooled (FC) magnetization susceptibility under very low magnetic field amplitudes. $\Delta T_c$ is of the order of 1.2 K for sample 1 (this will be then close to the intrinsic width of the superconducting transition in this compound), whereas it is 6.8 K for sample 2, which indicates the presence of important extrinsic $T_c$-inhomogeneities in this sample [8]. Taking into account that $\mu_0 H_{c2}(0)$ in $La_{1.9}Sr_{0.1}CuO_4$ is of the order of 30 T, one may crudely estimate from Eq.(1) (and Fig.1) that for sample 1 the magnetization measurements along the temperature isotherms may avoid the full inhomogeneous region by just using applied fields above 3 T.

As shown in Ref. 8, the *fine* behavior of the magnetization anomalies in the full inhomogeneous region strongly depends on the $T_c$ distribution characteristics. This conclusion applies particularly to the amplitude and temperature behavior of $H_{up}$, the magnetic field at which the differential magnetic susceptibility changes from negative to positive when the applied magnetic field increases: The increase with temperature of $H_{up}(T)$ observed in sample 2 may be attributed to the marked asymmetry of its $T_c$-distribution, in turn due to the combination of a large $\Delta T_c$ and the characteristic bell-shaped $T_c$ dependence on the doping level. To check these conclusions, in Fig.2 we compare the amplitude of the magnetization at $H_{up}(T)$ measured in the two samples studied in Ref. 8, together with the derivative of the field cooled susceptibility (normalized at their values at $\bar{T}_c$). These results further suggest that the anomalies of the magnetization isotherms in the low field regime, sometimes called *fragile London rigidity* [6], could be in fact directly related to $T_c$-inhomogeneities.

## 3. On the relationship between precursor diamagnetism and paraconductivity in the underdoped $La_{1.9}Sr_{0.1}CuO_4$.

The results on $\Delta\chi_{ab}$ of Ref. 8 are particularly useful to check the influence of the $T_c$-inhomogenities on $\Delta\chi_{ab}/T\Delta\sigma_{ab}$. As first stressed by Vidal and coworkers [13], on the grounds of the GGL approach the *direct* fluctuation contributions to $\Delta\chi_{ab}$ and $\Delta\sigma_{ab}$ are related by (in MKSA units),

$$\Delta\chi_{ab}(\varepsilon) \,/\, T\,\Delta\sigma_{ab}(\varepsilon) = 2.79 \times 10^5 \, \xi_{ab}^2(0), \qquad (2)$$

where $\varepsilon \equiv \ln(T/\bar{T}_c)$ is the reduced temperature. This relationship has been since then used in different works, in particular to further probe both its adequacy in optimally doped cuprates and the absence of *indirect* (e. g., Maki-Thompson) fluctuation contributions to the paraconductivity in these HTSC [13-15]. A similar relationship has been used recently in comparing the fluctuations effects on the high-frequency conductivity and $\Delta\chi_{ab}$ in underdoped cuprates [10, 12].

The data points in Fig. 3 correspond to $\Delta\chi_{ab}(\varepsilon)/T\Delta\sigma_{ab}(\varepsilon)$ obtained by using for $\Delta\chi_{ab}(\varepsilon)$ some of the measurements performed in samples 1 and 2 under a a 5 T applied magnetic field, and presented in Fig. 7 of Ref. 8. As we have already stressed, for sample 1 this field amplitude suffices to quench the full inhomogeneous region, which in this sample is near the expected intrinsic-like one [8]. In contrast, even under these field amplitudes the $\Delta\chi_{ab}(\varepsilon)$ data for sample 2 are still deeply affected by extrinsic inhomogeneities. For $\Delta\sigma_{ab}(\varepsilon)$ we have used the data of Fig. 11(a) of Currás *et al.* [9] [with $\bar{T}_c = T_{cI} - 0.6$ K, which leads to 2D fluctuations, in agreement with $\Delta\chi_{ab}(\varepsilon)$]. The solid line was obtained from Eq. (2), with $\xi_{ab}(0) \approx 3.0 \times 10^{-9}$ m, in good agreement with the values expected for this compound [3, 4].

The results of Fig.3 show that when the inhomogeneities are overcome (as for sample 1), also in the underdoped $La_{1.9}Sr_{0.1}CuO_4$ the relationship between these two observables agree at a quantitative level, in an extended temperature window above $\bar{T}_c$, with the GGL predictions [Eq.

(2)]. This last result confirms at a quantitative level the conclusion that could be obtained by just comparing our earlier measurements of $\Delta\chi_{ab}(\varepsilon)$ [16] and $\Delta\sigma_{ab}(\varepsilon)$ [9]. Complementarily, our present results also show the strong disagreement with the GGL prediction when the $\Delta\chi_{ab}(\varepsilon)$ data for sample 2 are used. For both samples, this disagreement will still increase by more than two orders of magnitude if the $\Delta\chi_{ab}(\varepsilon)$ data obtained under very low-field amplitudes are used (see Fig. 7 of Ref. 8). In addition, these spurious effects may be strongly enhanced by inadequate $\bar{T}_c(0)$ and background estimations when extracting $\Delta\chi_{ab}(\varepsilon)$ from magnetization data deeply affected by chemical disorder [3, 4].

## 4. Conclusions.

The results presented here further suggest that once the chemical disorder effects are overcome, both the in-plane precursor diamagnetism and the in-plane paraconductivity in the underdoped $La_{1.9}Sr_{0.1}CuO_4$ superconductor may be explained at a phenomenological level, simultaneous and consistently, in terms of the GGL approach. Our present results also suggest the usefulness of a check in terms of Tc inhomogeneities of the findings of Bilbro and coworkers [10], which could also be affected by chemical disorder, mainly through the independent magnetization measurements used in that work. This last conclusion seems to be supported by the phase diagram for the onset of superconducting correlations proposed in Ref. 11, which seems to agree qualitatively with the paraconductivity onset observed in the earlier measurements of Curras and coworkers in the same compounds [9], this last in turn accounted for by the so-called total energy cutoff [17].

**Acknowledgments.** This work was supported the Spanish MICINN and ERDF (FIS2010-19807), the Xunta de Galicia (2010/XA043 and 10TMT206012PR), and the european project ENERMAT.

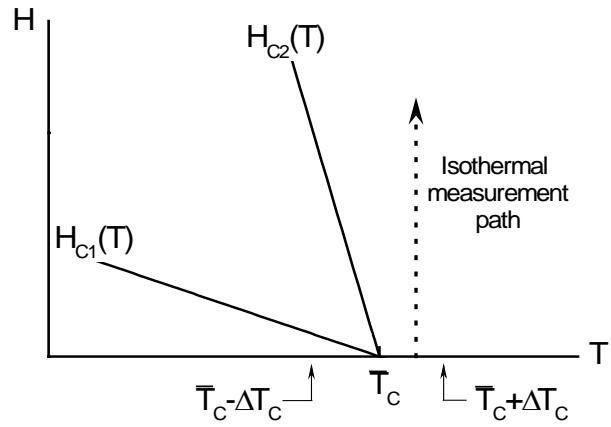

Fig.1. Schematic phase diagram for an inhomogeneous sample with a Gaussian $T_c$ distribution.

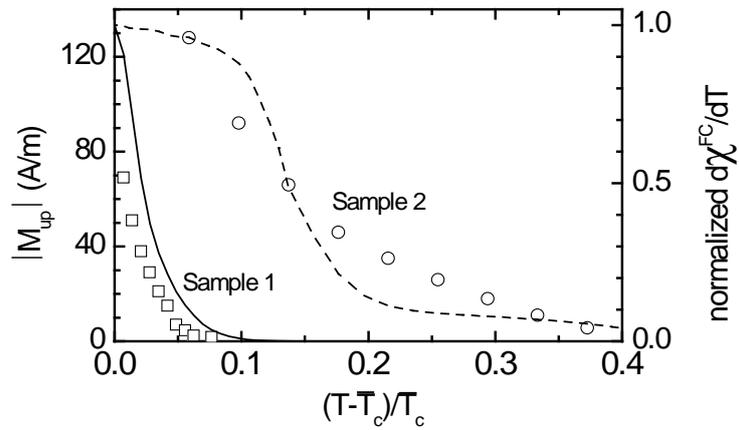

Fig. 2. Temperature dependence of the magnetization amplitude at $H_{up}$, the magnetic field at which the magnetic susceptibility isotherms change from decreasing to increasing when the applied magnetic field increases. The lines are the temperature derivative of the field-cooling susceptibility, normalized at their value at $\bar{T}_c$.

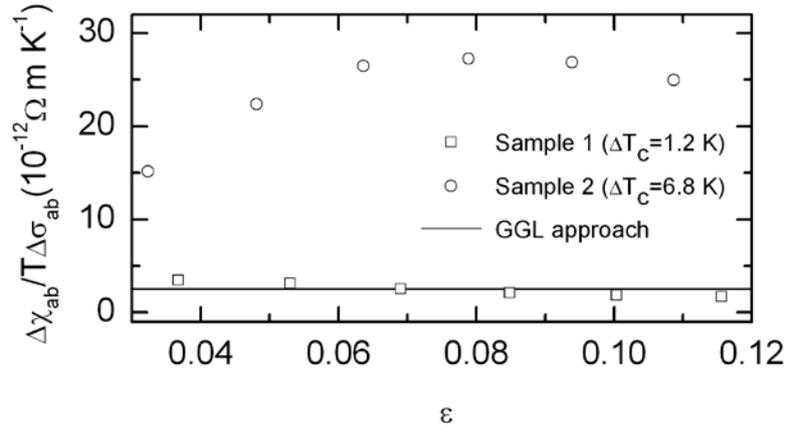

Fig. 3. Relationship between the precursor diamagnetism of the two $La_{1.9}Sr_{0.1}CuO_4$ samples studied in Ref. 7 and the paraconductivity, reported in Ref. 15, in a film of the same composition. In these examples the applied magnetic field was 5 T, which for sample 1 is strong enough to quench its full inhomogeneous region. The line is the GGL prediction [Eq. (2)].